\documentclass[12pt]{article}
\usepackage{amsmath}
\usepackage{graphicx}
\begin{document}

\vspace*{2cm}
\begin{center}
 \Huge\bf
Topos-theoretic Model \\ of the Deutsch multiverse
\vspace*{0.25in}

\large

Alexander K.\ Guts
\vspace*{0.15in}

\normalsize

Department of Computer Science, Omsk State University \\
644077 Omsk-77 RUSSIA
\\
\vspace*{0.5cm}
E-mail: guts@univer.omsk.su  \\
\vspace*{0.5cm}
November 21, 2001\\
\vspace{.5in}
ABSTRACT
\begin{quote}
The Deutsch multiverse is collection of parallel universes.
In this article a formal theory and a topos-theoretic model of the Deutsch
multiverse are given. For this the Lawvere-Kock Synthetic 
Differential Geometry
and topos  models for smooth infinitesimal analysis are used.
Physical properties of multi-variant and many-dimensional parallel
universes are discussed. Quantum fluctuations of universe 
geometry are considered.
Photon ghosts in parallel universes are found.
\end{quote}
\end{center}

\vfill
${}$

This paper was published in Russian journal "Mathematical Structures
 and Modeling", No.8,
76-90 (2001) (ftp://cmm.univer.omsk.su/pub/sbornik8/guts.zip).
\newpage


\def\R{{{\rm I} \! {\rm R}}}
\def\C{{{\rm I} \! \hspace*{-1.3mm} {\rm C}}}
\def\L{{{\rm I} \! {\rm L} }}
\def\D{{{\cal D} }}
\def\M{{{\cal M} }}
\def\F{{{\cal F} }}
\def\G{{{\cal G} }}
\def\Z{{{\cal Z} }}
\def\T{{{\cal T} }}
\def\d{{{\rm \Delta} \!\!\!\! {\Delta}}}
\def\N{{{\rm I} \! {\rm N} }}

\section*{Introduction}

In David Deutsch 's book \cite{1} the sketch of structure of  physical
reality  named Multiverse which is set of the parallel universes is given.
Correct description of the Multiverse
(as Deutsch considers) can be done
only within the framework of the quantum theory.

In this article a formal theory and a topos-theoretic model of the Deutsch
multiverse are given.

We wish to preserve the framework of the mathematical apparatus
of the 4-dimensional General theory of the relativity,
and so we shall consider the Universe as concrete 4-dimensional
Lorentz manifold $<R^{(4)}, g^{(4)}>$ (named space-time).

Our purpose is to give an opportunity to take into account presence
parallel universes, i.e. other universes being most various 4-dimensional
pseudo-Riemannian manifolds which are belonged to
special hyperspace of any dimension.

Moreover, hyperspaces should be as is wished much; the geometry,
topology, dimension of hyperspaces should be as much as various
that always it was possible to find uncountable number of the
universes as much as similar to ours, and simultaneously should
exist as is wished a lot of installed, completely unlike the world
in which we live.

The structure of a physical reality should take into account whim
of a conceiving essence to see it in every possible conceivable
forms, having thus rather poor research toolkit which basis should be
the theory of a relativity and the quantum mechanics.

We are not going to pass to
 many-dimensional theories such as Kaluza-Klein theory. No.
We emphasize that a basis of the Multiverse theory should be
the 4-dimensional metric $g^{(4)}$.

\section{Formal theory of Multiverse}

We create the theory of Multiverse as formal theory $\T$ which
is maximally similar to
 the General theory of Relativity, i.e. as theory of {\it one}
4-dimensional universe,
 but other parallel universes must
appear under costruction of models of formal theory.

The basis of our formal theory  $\T$ is the Kock-Lawvere
Synthetic Differential Geometry (SDG) \cite{Cock}.

SDG has not any set-theoretic model because Kock-Lawvere axiom
is  incompatible with Law of  excluded middle.
Hence we shall construct formal theory of Multiverse
with intuitionistic logic. Models for this theory
are topos-theoretic models.

In SDG the commutative ring $R$ is used instead of real field
 $\R$. The ring $R$ must satisfy the following axioms
 \footnote{We give some axioms. Other axioms see in \cite[Ch.VII]{Moer}.}:

{\small
\begin{itemize}
  \item [(A1)] $<R,+,\cdot,0,1>$ is commutative ring.
  \item [(A2)] $R$ is local ring, i.e.
  $$
  \begin{array}{l}
    0=1 \Longrightarrow \bot \\
    \exists y\ (x\cdot y=1) \exists y \ (1-x)\cdot y=1.
  \end{array}
  $$
  \item [(A3)] $<R,<\ >$ is real Euclidean ordered local ring, i.e.
  $<$ is transitive relation such that
  $$
\begin{array}{l}
(a)\ 0<1,\ (0<x\ \& \ 0<y \Longrightarrow 0<x+y\ \& \ 0<x\cdot y), \\
(b)\ \exists y (x\cdot y=1)\Longleftrightarrow (0<x \vee x<0), \\
(c)\ 0<x \Longrightarrow \exists y(x=y^2)\ (Euclidean\  property).
\end{array}
$$
  \item [(A4)] $\leq$\ is a preorder, i.e. reflexive and transitive relation,
and
$$
\begin{array}{l}
  (a)\ 0\leq 1,\ (0\leq x\ \& \ 0\leq y \Longrightarrow 0\leq x+y\ \& \
   0\leq x\cdot y),\ 0\leq x^2, \\

   (b)\ (x\  \mbox{is nilpotent, i.e.}\ x^n=0)   \Longrightarrow 0\leq x.
   \end{array}
$$
  \item [(A5)] $<$\ \mbox{and}\ $\leq$ are compactible in the following sence:
$$
\begin{array}{l}
  (a)\ x<y \Longrightarrow x \leq y,\\
  (b)\ x<y\ \& \ y\leq x \Longrightarrow \bot.
\end{array}
$$
\item [(A6)] (Kock-Lawvere axiom). Let $D=\{x\in R : x^2=0\}$. Then
  $$
  \forall(f\in R^D)  \exists ! (a,b)\in R\times R\ \forall 
d\in D( f(d)=a+b\cdot d).
  $$

\smallskip

  \item [(A7)] (Integration axiom).
  $$
  \forall f\in R^{[0.1]}\exists!g\in
  R^{[0.1]}(g(0)=0 \ \& \ \forall x\in [0,1]\ (g^\prime (x)=f(x)),
  $$

where $[0,1]=\{x\in R : 0\leq x\ \& \ x\leq 1\}$ and $g^\prime(x)$ is the only
$b$ such that $\forall d\in D(g(x+d)=g(x)+b\cdot d)$.

  We use the symbolic  record:
  $$
  g(x)=\int\limits_0^1f(t)dt.
  $$
\item [(A8)]   ${\displaystyle \forall x\in [0,1]\ ( 0<f(x) \Longrightarrow
  0<\int\limits_0^1f(x)dx})$.
  \item [(A8$^\prime$)] ${\displaystyle \forall x\in [0,1]\ ( 0\leq f(x)
 \Longrightarrow
  0\leq\int\limits_0^1f(x)dx})$.
\item [(A9)] (Inverse function theorem).
$$
\forall f\in R^R \forall x\in R (f^\prime(x)
 \mbox{\ inversible}\Longrightarrow
$$
$$ \Longrightarrow
\exists \mbox{\ open}\ U,V(x\in U\ \&\ f(x)\in V\ \&\ f|_U\to V\  
\mbox{is a bijection})).
$$
\item [(A10)] $N\subset R$,\ \mbox{i.e.}\ $\forall x\in N\ 
\exists y\in R (x=y)$.
\item [(A11)] $R$ is Archimedean for $N$, i.e. $\forall x\in R\ 
\exists n\in N (x<n)$.
\item [(A12)] (Peano axioms).
$$
\begin{array}{l}
  0\in N  \\
  \forall x\in R\ (x\in N \Longrightarrow  x+1\in N)\\
\forall x\in R\ (x\in N\ \&\ x+1=0 \Longrightarrow  \bot).
\end{array}
$$
\end{itemize}
}

Ring  $R$ includes real numbers from $\R$ and has new elements named
{\it infinitesimals} belonging to "sets"
$$
D=\{d\in R : d^2=0\}, ..., D_k=\{d\in R : d^{k+1}=0\},...,
$$
$$
\d=\{x\in R : f(x)=0, \ \mbox{all}\ f\in m_0^g \},
$$
where $m^g_{\{0\}}$ is ideal of functions having zero germ  at 0, 
i.e. vanishing
in a neighbourhood of 0.
причем

We have
$$
D\subset D_2\subset ... \subset D_k\subset... \subset \d.
$$

For given system of axioms we can construct \cite{gu,gugri} 
Riemmanian geometry
for four-dimensional (formal) manifolds $<R^4, g^{(4)}>$. These 
manifolds are basis
for the Einstein theory of gravitation.

\smallskip
We postulate that {\it multiverse is four-dimensional space-time in SDG,
i.e. is a formal Lorentz manifold $<R^4, g^{(4)}>$ for which
the Einstein field equations are held:
\begin{equation}\label{ein}
R^{(4)}_{ik}-\frac{1}{2}g^{(4)}_{ik}(R^{(4)}-2\Lambda)=
\frac{8\pi G}{c^4}T_{ik}.
\end{equation}
}

\smallskip
A solution of these equations is 4-metric $g^{(4)}$.

\smallskip

Below we consider the physical consequences of our theory in so 
called well-adapted
models of the form ${\bf Set}^{\L^{op}}$ which contain as full subcategory
the category of smooth manifolds $\M$.

\section{Smooth topos models of multiverse}

Let  $\L$ be  dual category for category of finitely generated 
$C^\infty$-rings.
It is called {\it category of loci} \cite{Moer}. The objects of  
$\L$ are €вл€ютс€
finitely generated $C^\infty$-rings, and morphisms are reversed morphisms of
category of finitely generated $C^\infty$-rings.

The object (locus) of $\L$ is denoted as $\ell A$, where  $A$ is 
a $C^\infty$-ring.
Hence, $\L$-morphism $\ell A \to \ell B$ is $C^\infty$-homomorphism $B \to A$.

A finitely generated $C^\infty$-ring $\ell A$ is isomorphic to ring of the
form $C^\infty(\R^n)/I$ (for some natural number $n$ and
some некоторого finitely generated function ideal~$I$).

Category  ${\bf Set}^{\L^{op}}$ is topos. We consider topos
${\bf Set}^{\L^{op}}$ as model of formal theory of multiverse. Only some 
from axioms
(A1)-(A12) are true in topos model ${\bf Set}^{\L^{op}}$  \footnote{
One can take as models topoi $\F, \G$ и $\Z$ and others
\cite[Appendix 2]{Moer}. All axioms
(A1)-(A12) are true for these topoi (see \cite[p.300]{Moer})}.

\smallskip
With the Deutsch point of view  the transition to concrete model
of formal theory is creation of {\it virtual reality} \footnote{This  thought
belong to Artem Zvaygintsev.}. Physical Reality that we  perceive was called
by Deutsch {\it Multiverse}
 \footnote{Multiverse = many (multi-) worlds;
 universe is one (uni) world. }. Physical Reality is also
virtual reality which was created our brain
\cite[p.140]{1}.

\medskip
A model of multiverse is {\it generator of virtual reality} which
has some {\it repertoire of  environments}.
Generator of virtual reality creates environments and we
observe them.  Explain it.

\medskip
Under interpretation $i: {\bf Set}^{\L^{op}}\models \T$ of formal
multiverse theory $\T$ in topos ${\bf Set^{\L^{op}} }$
the objects of theory, for example, ring $R$, power $R^R$ and so on are 
interpreted
as objects of topos, i.e.
functors $F=i(R)$, $F^F=i(R^R)$ and so on. Maps, for example,
$R\to R$,  $R\to R^R$ are now morphisms of topos
 ${\bf Set}^{\L^{op}}$, i.e. natural transformations of functors:
$F\to F$, $F\to F^F$.

Finelly, under interpretation of language of formal multiverse theory
we must interpret elements of ring
$R$ as "elements" of functors $F\in {\bf Set}^{\L^{op}}$.   In other words
we must give interpretation for relation $r\in R$. It is very
difficult task because functor $F$ is defined on category of loci $\L$;
its independent variable is arbitrary locus $\ell A$, and
dependent variable is a set $F(\ell A)\in {\bf Set}$. To solve this problem
we consider {\it generalized elements} $x\in_{\ell A}F$ of functor $F$.

Generalized element $x\in_{\ell A}F$, or {\it element  $x$ of functor 
 $F$ at stage
${\ell A}$}, is called element $x\in F({\ell A})$.

Now we element $r\in R$ interpret as generalized element $i(r)\in_{\ell A}F$.
We have such elements so much how much loci. Transition to model
${\bf Set}^{\L^{op}}$  causes "reproduction" of element $r$. It begins
to exist in  infinite number of variants $\{i(r): i(r)\in_{\ell A}F, 
\ell A\in \L \}$.

Note that since  4-metric $g^{(4)}$ is element of object
$R^{R^4\times R^4}$ then "intuitionistic" 4-metric
begins
to exist in  infinite number of variants $i(g)^{(4)}\in_{\ell A}
i(R^{R^4\times R^4})$.
Denote such variant as $i(g)^{(4)}(\ell A)$.

For simplification of interpretation we shall operate with objects of models
${\bf Set}^{\L^{op}}$. In other words, we shall write
$g^{(4)}(\ell A)$ instead of
$i(g)^{(4)}(\ell A)$.

Every variant $g^{(4)}(\ell A)$
of 4-metric $g^{(4)}$ satisfies to "own" Einstein equations~\cite{gu}
\begin{equation}\label{einA}
R^{(4)}_{ik}(\ell A)-\frac{1}{2}g^{(4)}_{ik}(\ell A)[R^{(4)}(\ell
A)-2\Lambda(\ell A)]=\frac{8\pi G}{c^4}T_{ik}(\ell A).
\end{equation}
(Constants $c,G$ can have different values for different stages $\ell A$).

\begin{figure}[h]
\centering\includegraphics[width=13.3cm]{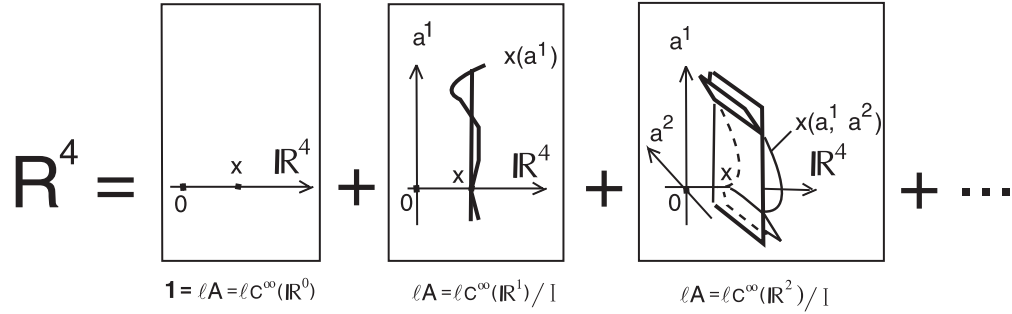}
\caption{\footnotesize  Physical (virtual) Reality $R^4$ as sum of 
many-dimensional
hyperspaces (environments) $R^4(\ell A)$. Every hyperspace contains 
a foliation
 which consists of parallel 4~-~dimensional
universes ($a=const$).}
\label{gu11}
\end{figure}

Previously before going any farther, we shall point to existance of 
Yoneda embedding
$$
y:\L\hookrightarrow {\bf Set}^{\L^{op}},
$$
$$
y(\ell A)=Hom_\L(-, \ell A).
$$
Assume that ring $R$ is interpreted as
functor $y(\ell C^\infty(\R))$, i.e. $i(R)=y(\ell C^\infty(\R))$.
Write $\ell A$ instead of $y(\ell A)$ and omit symbol $i$. Then we have
$$
R(-)=\ell C^\infty(\R)(-)=Hom_\L(-, \ell C^\infty(\R)).
$$
Similarly
$$
R^{R^4\times R^4}(\ell A)=Hom_\L(\ell A, R^{R^4\times R^4})=
Hom_\L(\ell A\times (R^4\times R^4), R)=
$$
$$
=Hom_\L(\ell C^\infty(\R^m)/I\times \ell C^\infty(\R^4)\times 
\ell C^\infty(\R^4),\ell
C^\infty(\R))=
$$
$$
=Hom_{\L^{op}}(\ell C^\infty(\R),
C^\infty(\R^m)/I\otimes_\infty C^\infty(\R^4)\otimes_\infty C^\infty(\R^4))=
$$
$$
=Hom_{\L^{op}}(C^\infty(\R), C^\infty(\R^{m+8})/(I,\{0\}))=
$$
$$
=Hom_\L(\ell C^\infty(\R^{m+8})/(I,\{0\}),\ell C^\infty(\R)),
$$
where $\ell A=\ell C^\infty(\R^m)/I$, $\otimes_\infty$ is symbol 
of coproduction of
$C^\infty$-rings and  under calculation the folowing formulas are used:
$$
C^\infty(\R^n)\otimes_\infty C^\infty(\R^k)=C^\infty(\R^{n+k}),
$$
$$
\frac{\ell A\to \ell C^{\ell B}}{\ell B\times \ell A\to \ell C}.
$$
It follows from this that when $\ell A=\ell C^\infty(\R^m)$ then
$$
g^{(4)}(\ell A)=[g\in_{\ell A}R^{R^4\times R^4}]\equiv
g^{(4)}_{ik}(x^0,...,x^3, a)dx^idx^k, \ \ a=(a^1,...,a^m)\in \R^m.
$$
Four-dimensional metric $g^{(4)}_{ik}(x^0,...,x^3, a)$ we extend to 
(4+m)-metric
in space~$\R^{4+m}$
\begin{equation}\label{met}
 g^{(4)}_{ik}(x^0,...,x^3, a)dx^idx^k-da^{1^2}-...-da^{m^2}.
\end{equation}
We get  $(4+m)$-dimensional geometry.

Symbolically procedure of creation of many-dimensional variants of geometry
by means of intuitionistic 4-geometry $g^{(4)}$
 one can represent in the form of formal
sum
$$
g^{(4)}=c_0\cdot\hspace*{-3mm}\underbrace{[g^{(4)}\in_{\bf 1}R^{R^4
\times R^4}]}_{\mbox{\footnotesize 4-geometry}}+
c_1\cdot\underbrace{[g^{(4)}\in_{\ell C^\infty(\R^1)}R^{R^4\times
R^4}]}_{\mbox{\footnotesize 5-geometry}}+...
$$
$$
...+c_{n-4}\cdot\underbrace{[g^{(4)}\in_{\ell C^\infty(\R^{n-4})}
R^{R^4\times R^4}]}_{\mbox{\footnotesize n-geometry}}+...,
$$
where coefficients $c_m$ are taked from the field of complex numbers.

Because number of stages is infinite,   we must
write integral instead of sum:
\begin{equation}\label{metr1}
  g^{(4)}=\int\limits_{\L}\D[\ell A]c(\ell A)[g^{(4)}
  \in_{\ell C^\infty(\R^{n-4})}R^{R^4\times
R^4}].
\end{equation}
Use denotations of quantum mechanics \footnote{Dirac denotations:
 $|P\rangle=\psi(\xi)\rangle\equiv \psi(\xi)$; in given case $\psi(\xi)$ is
$g^{(4)}$ (representative
 of state $|P\rangle$), and
 $|P\rangle$ is $|g^{(4)}\rangle$ \cite[p.111-112]{Dir}.}:
$$
g^{(4)}\to |g^{(4)}\rangle, \ \ \
[g^{(4)}\in_{\ell C^\infty(\R^{n-4})}R^{R^4\times R^4}]\to 
|g^{(4)}(\ell A)\rangle.
$$
Then (\ref{metr1}) is rewrited in the form
\begin{equation}\label{metr2}
|g^{(4)}\rangle=\int\limits_{\L}\D[\ell A]c(\ell A)|g^{(4)}(\ell A)\rangle.
\end{equation}
Consequently, formal the Kock-Lawvere 4-geometry $<R^4,g^{(4)}>$ is 
infinite sum of
 of classical many-dimensional pseudo-Riemmanian geometries which
contain foliation of 4-dimensional parallel universes (leaves) (under 
fixing $a=const$).
Geometrical properties of these universes as it was shown in \cite{guz,guz1}
to be different even within the framework of one stage $\ell A$.
About nature of coefficients  $c(\ell A)$
we say below in \S 5.

\medskip
Now we  recall about environments of virtual reality
 which must appear under referencing to model of multiverse, in this instance,
 to model ${\bf Set}^{\L^{op}}$. This model is generator of virtual reality.
It is not difficult to understand that generalised element
$|g^{(4)}(\ell A)\rangle$ is  metric of concrete  environment 
(=hyperspace $R^4(\ell A)$)
 with "number" $\ell A$.
 In other words, study of any object of theory $\T$ at stage $\ell A$
 is transition to one of the environments
 from repertoire of virtual reality generator ${\bf Set}^{\L^{op}}$.

\section{The Deutsch-G\"odel Multiverse}

As example of multiverse we consider
cosmological solution of Kurt G\"odel~\cite{ged}
\begin{equation}\label{mged}
g^{(4)}_{ik}=\alpha^2
\left(\begin{array}{cccc}
  1 & 0 & e^{x^1} & 0 \\
  0 & -1 & 0 & 0 \\
  e^{x^1} & 0 & e^{2x^1}/2 & 0 \\
  0 & 0 & 0 & -1
\end{array}\right).
\end{equation}
This metric satisfies the Einstein equations (\ref{ein}) with energy-momentum
tensor of dust matter
$$
T_{ik}=c^2\rho u_iu_k,
$$
if
\begin{equation}\label{gcond}
\frac{1}{\alpha^2}=\frac{8\pi G}{c^2}\rho, \ \ \Lambda=-\frac{1}{2\alpha^2}=
-\frac{4\pi G\rho}{c^2}.
\end{equation}
Take
\begin{equation}\label{const}
\alpha=\alpha_0+d,\ \Lambda=\Lambda_0+\lambda,\ \rho=\rho_0+\varrho,
\end{equation}
where $d,\lambda,\varrho\in D$ are infinitesimals and substitute these
in (\ref{gcond}). We get
$$
\frac{1}{(\alpha_0+d)^2}=\frac{1}{\alpha_0^2}-\frac{2d}{\alpha_0^3}=
\frac{8\pi G}{c^2}(\rho_0+\varrho),
$$
$$
2\Lambda_0+2\lambda=-\frac{1}{\alpha_0^2}+\frac{2d}{\alpha_0^3},\ \ 
\Lambda_0+\lambda=
-\frac{4\pi G\rho_0}{c^2}-\frac{4\pi G\varrho}{c^2}.
$$
Suppose that $\alpha_0,\Lambda_0,\rho_0\in \R$ are satisfied to 
relations (\ref{gcond}). Then
$$
\lambda=-\frac{4\pi G}{c^2}\varrho, \ \ 
d=-\frac{4\pi G\alpha_0^3}{c^2}\varrho.
$$
Under interpretation in smooth topos ${\bf Set}^{\L^{op}}$ infinitesimal
 $\varrho\in D$ at stage
$\ell A=C^\infty(\R^m)/I$ is class of smooth functions of the 
form $\varrho(a)mod\ I$,
where
$[\varrho(a)]^2$ $ \in I$ \cite[p.77]{Moer}.

Consider the properties  of the Deutsch-G\"odel multiverse at stage
 $\ell A=$ $\ell C^\infty(\R)/(a^4)$ \footnote{As $(f_1,...,f_k)$ is denoted
 ideal of ring
$C^\infty(\R^n)$ generated dy functions $f_1,...,f_k\in C^\infty(\R^n)$, 
i.e. having the form
$\sum_{i=1}^kg_if_i$, where  $g_1,...,g_k\in C^\infty(\R^n)$ are arbitrary
smooth functions.}, where  $a\in\R$. Obviously that it is possible to take 
infinitesimal
 of form  $\varrho(a)=a^2$. Multiverse at this stage is 5-dimensional
hyperspace. This hyperspace contains a foliation, leaves of which are defined
by the equation $a=const$.
 The leaves are parallel universes in hyperspace (environment) $R^4(\ell A)$
with metric $g^{(4)}(\ell A)=g^{(4)}_{ik}(x,a)$ defined formulas 
(\ref{mged}), (\ref{const}).
Density of dust matter $\rho=\rho_0+\varrho(a)$ grows from classical
value $\rho_0\sim 2\cdot 10^{-31}$ {\it g/cm$^3$} to $+\infty$
under $a\to \pm\infty$. Cosmological constant  grows  also infinitely 
to $-\infty$.
Hence parallel universes  have different from our Universe
physical properties.

At stage $\ell A=\ell C^\infty(\R)/(a^2)$ \ $\varrho(a)=a$ and 
$\rho=\rho_0+\varrho(a)\to
-\infty$
under $a\to -\infty$, i.e. $\rho$ is not phisically interpreted (we 
have "exotic" matter
with negative density).

Finally, at stage ${\bf 1}=\ell C^\infty(\R)/(a)$ all 
$\varrho(a)=d(a)=\lambda(a)=0$,
i.e. we have classical the G\"odel universe.

\section{Quantum properties of parallel\\ universe geometry}

We apply the ideas of the
Wheeler quantum geometrodynamics to our formal theory of
multiverse. So, formula for probability amplitude of transition 
from 3-geometry $g^{(3)}$
of physical 3-space to 3-geometry $h^{(3)}$ has the form of "double" 
Feinman integral
over 4-dimensional trajectories $g^{(4)}$:
$$
\langle g^{(3)}|h^{(3)}\rangle=\int\limits_{\L}\D[\ell
A]\int\limits_{ g^{(3)}(\ell A) }^{ h^{(3)}(\ell A) } 
\D[g^{(4)}(\ell A)]e^ { \frac{
i}{\hbar} S[g^{(4)}(\ell A)] }, $$ where $$ S[g^{(4)}(\ell A)]=
\kappa_m(\ell
A)\int\limits_{\R^{4+m}} \sqrt{ -det||g^{(4)}(\ell A)|| } 
R^{(4)}(\ell A)d^4xda^m
$$
 is
action in space $<\R^{4+m},g^{(4)}(\ell A)>$.

We see that this Feinman integral over trajectories $g^{(4)}$ is
infinite number of integrals over $(4+m)$-dimensional trajectories  
$g^{(4)}(\ell A)$
of the form~(\ref{met}).

We can found quantum fluctuations of 4-metric
$g^{(4)}\to g^{(4)}+\Delta g^{(4)}$ which do not give any distortion 
in interference
picture.

Assume that $det||g^{(4)}(\ell A)||\sim~1$. Then we get for fluctuations
in $(4+m)$-dimensional domain with sizes  $L^4\times L_1^m$:
\begin{equation}\label{fluc}
\Delta g^{(4)}(\ell A)\sim \frac{L^*}{L}\left(\frac{T}
{L_1}\right)^{\frac{m}{2}},
\end{equation}
where
$$
L^*=\sqrt{\frac{G\hbar}{c^3}}\sim 10^{-33}\mbox{{\it cm}}
$$
is Planck length. Here $\kappa_m(\ell A)\sim c^3/(\hbar GT^m)$, where
$T$ [{\it cm}] is value characterizing "size" of additional dimensions.

It follows from (\ref{fluc}) that under  $L\sim L^*, L_1\sim T$ all 
fluctuations
$\Delta g^{(4)}(\ell A)\sim~1$, i.e. geometry and topology froth.

As it is shown in  \cite{guz4,guz5} fluctuations can take a place
at large scale of space and time. Here the main role belongs to
additional dimensions which are appeared under consideration
of multiverse state at different stages $\ell A$.

\section{Electrons-twins}

Deutsch has expected that parallel universe is formed from {\it shadow}
elementary particles accompanying each {\it real}\  particle. The 
real particles
we can see or find by means of instruments, but the shadow particles 
are invisible.
They can be found only through their influence with real particles 
\cite[p.48]{1}.
"Between real and shadow photons does not exist any differences:
each photon is  perceived in one universe and is not  perceived in 
all other parallel
universes".

The Dirac equation in SDG
\begin{equation}\label{dir}
{\displaystyle  i\hbar {\gamma}^{(k)}\frac{\partial\psi}
{\partial x^{\scriptscriptstyle k}}-mc\psi =0},
\end{equation}
for Minkowsky space-time, i.e. in the Deutsch-Minkowsky multiverse 
$M^4$ with metric
\begin{equation}\label{1}
 ds^2=dx^{0^2}-dx^{1^2}- dx^{2^2}-dx^{3^2},
\end{equation}
has, for example, the following solution
\begin{equation}\label{psi}
\psi(x)=\left(\begin{array}{c}
\  \ 1 \\
\  \ 1 \\
  -1 \\
 \ \ 1
\end{array}\right)e^{\vspace*{-1cm}\frac{mc}{\hbar}x^2+g(x^3+x^0)+i
\theta\cdot
f(x^3+x^0)}.
\end{equation}
This solution under $\theta\cdot f(x^3-x^0)=const$ is spinor ghost  
\footnote{This solution
was found by Elena Palesheva.}, i.e.
has zero energy-momentum tensor of field $\psi(x)$:
\begin{equation}\label{4}
\begin{split}
T_{ik}=\frac{i\hbar c}{4}\left\{{\psi}^*\gamma^{(0)}{\gamma}_{(i)}
\frac{\partial\psi}{\partial
x^{\scriptscriptstyle k}}-\frac{\partial {\psi}^*}{\partial x^{
\scriptscriptstyle k}}{\gamma}^{(0)}\gamma_{(i)}\psi+\right.\hspace*{4cm}\\
\left.\hspace*{4cm}+{\psi}
^*{\gamma}^{(0)}{\gamma}_{(k)}\frac{\partial\psi}{\partial 
x^{\scriptscriptstyle i}}
-\frac{\partial {\psi}^*}{\partial x^{\scriptscriptstyle
i}}{\gamma}^{(0)}\gamma_{(k)}\psi\right\}.
\end{split}
\end{equation}
Hence, spinor ghost $\psi$ does not possess neither energy, nor 
momentum. So they
can not be fixed any instrument.
E.V.~Palesheva has offerred \cite{Pal} to identify the spinor ghosts
 with the Deutsch shadow particles.

Solution $\psi$ is connected \footnote{See note 5.} with Dirac ket-vector
$|\Psi\rangle$  represented in the
form of sum \footnote{The given formula
has relation to the Everett interpretation of quantum mechanics~\cite{ever}. }
\begin{equation}\label{sol1}
|\Psi\rangle=\int\limits_{\L}\D[\ell A]a(\ell A)|\Psi(\ell A)\rangle.
\end{equation}
We interpret  $\psi=|\Psi\rangle$. Then $\psi^*\psi=\langle\Psi|\Psi
\rangle$ is
probability amplitude of electron and
\begin{equation}\label{sol2}
\int\limits_{R^4}\psi^*\psi d^4x=\int\limits_{R^4}\langle\Psi|
\Psi\rangle d^4x=1.
\end{equation}
Let
$$
\langle\Psi|=\int\limits_{\L}\D[\ell B]a^*(\ell B)\langle\Psi(\ell B)|.
$$
So
$$
1=\int\limits_{R^4}\langle\Psi|\Psi\rangle d^4x=
\int\limits_{\R^4} d^4x\int\limits_{\L}\D[\ell B]
\int\limits_{\L}\D[\ell A]a^*(\ell B)a(\ell A)
\langle\Psi(\ell B)|\Psi(\ell A)\rangle=
$$
$$
=\int\limits_{\L}\D[\ell B]a^*(\ell B)\int\limits_{\L}\D[\ell A]a(\ell A)
\left(\hspace*{2mm}\int\limits_{\R^4} d^4x\langle\Psi(\ell B)|
\Psi(\ell A)\rangle\right)=
$$
$$
=\int\limits_{\L}\D[\ell B]a^*(\ell B)\int\limits_{\L}\D[\ell A]a(\ell A)
\delta(\ell B-\ell A)
=\int\limits_{\L}\D[\ell B]a^*(\ell B)a(\ell B),
$$
where we take (as logical extension of  equality (\ref{sol2})) that
$$
\int\limits_{\R^4} d^4x\langle\Psi(\ell B)|\Psi(\ell A)\rangle=
\delta(\ell B-\ell A),
$$
$$
\int\limits_{\L} \D[\ell B]f(\ell B)\delta(\ell B-\ell A)=f(\ell A).
$$
Hence
$$
\int\limits_{\L}\D[\ell A]a^*(\ell A)a(\ell A)=1
$$
and we can assume that $a^*(\ell A)a(\ell A)$ is probability amplitude
of stage $\ell A$ characterizing probability of observation of electron
 at stage $\ell A$ of multiverse $M^4$.

\medskip
Such conclusion one allows to interpret $c^*(\ell A)c(\ell A)$, 
where $c(\ell A)$ is
complex coefficient in decomposition (\ref{metr2}) of 4-metric of
multiverse $<R^4, g^{(4)}>$, as probability (more exactly, amplitude 
of probability)
that multiverse is inhered in state $|g^{(4)}(\ell A)\rangle$
\footnote{Metric is gravitational field  defining geometry and in 
some sense topology
of space-time. So it is naturally to identify the state (the  environment)
 $|R^4(\ell A)\rangle$ of
multiverse  at stage $\ell A$ (see, for instance, pic.\ref{gu11})
 with state  $|g^{(4)}(\ell A)\rangle$ of 4-metric $g^{(4)}$.}.
\medskip

Take in  (\ref{psi}) number $\theta=1-\varepsilon$, where
$\varepsilon$
infinitesimal, i.e. $\epsilon\in \d=\{x\in~R | f(x)=0, \ 
\mbox{all}\ f\in m^g_{\{0\}}\}$, \
$m^g_{\{0\}}$ is ideal of functions having zero germ at 0.

If $\epsilon\in \d$ then $\varepsilon$ at stage $\ell C^\infty (\R^n)/I$
is defined by function $\varepsilon(a), a\in \R^n$ such that for any
$\phi\in m^g_{\{0\}}$ \ \
$\phi(\varepsilon(a))\in I$ \ \cite[p.77]{Moer}.

We have
$$
\phi(\varepsilon(a))=\phi(\varepsilon(0))+\sum_{|\alpha|=1}^\infty
\frac{1}{\alpha !}
 D^\alpha(\phi\circ\varepsilon)(0)a^\alpha=
$$
\begin{equation}\label{pe}
= \phi(\varepsilon(0))+
 \sum_{|\alpha|=1}^\infty\frac{1}{\alpha !}
 \left(\sum_{|\beta|=1}^{|\alpha|}
 D^\beta \phi(\varepsilon(0))P_\beta(\varepsilon(0))\right)a^\alpha,
\end{equation}
where $\alpha,\beta$ are multi-indexes and $P_\beta$ are some polynomials.

At stage $\ell C^\infty (\R^n)$ \ \ $\phi(\varepsilon(a))\in I=\{0\}$ for all
$\phi\in m^g_{\{0\}}$. So it follows from
(\ref{pe}) that $\phi(\varepsilon(0))=0$, and
$\varepsilon(0)=0$. Moreover
$$
\sum_{|\beta|=1}^{|\alpha|}
 D^\beta \phi(\varepsilon(0))P_\beta(\varepsilon(0))=0.
$$
But for any $\phi\in m^g_{\{0\}}$ \ \ $D^\beta \phi(0)=0$. Hence 
$\epsilon(a)$
is arbitrary function satisfing the condition $\varepsilon(0)=0$.

For field (\ref{psi}) we take that $\theta(a)=1-\varepsilon(a)$, where
$$
\varepsilon(0)=0, \ \ \varepsilon(a) > 0 \ \mbox{under}
 \ a\neq 0, \ \mbox{and}\ \epsilon(a)= 1\ \mbox{under}\ ||a||\geq r_0,
$$
and $f$ is some non-zero function. Then
we have at stage    $\ell A=\ell C^\infty (\R^n)$:
$$
\theta(a)=1-\varepsilon(a)=\left\{
\begin{array}{c}
  0\ \ \ \ \mbox{under}\ ||a||\geq r_0, \\
  >0\ \mbox{under}\ ||a|| < r_0.
\end{array}\right.
$$
Hence at stage    $\ell A=\ell C^\infty (\R^n)$ field $\psi$ is not 
spinor ghost
in our Universe ($a=0$) and in all universes
with $||a||<r_0$, but is ghost in papallel universes for which
 $||a||\geq r_0$. We can take number $r_0$ 
 so small that universes  "labeled" by parameter $a$ with $||a||<r_0$
 must be considered
as one universe due to quantum foam of topologies and geometries
($r_0$ is "thickness" of universe).
 This means that field $\psi$ is real particle in our Universe and
 shadow particle-twin in all other
 universes.

If we take $\theta\in \d$ such that
 $$
\theta(a) > 0 \ \mbox{under}  \ ||a-a_0||<r_0 \ \mbox{and}\
\theta(a)= 0\ \mbox{under}\ ||a||> r_0,
 $$
where $a_0\neq 0$ and $r_0<||a_0||$ then field $\psi$ at stage  
$\ell C^\infty (\R^n)$
is not spinor ghost in the universe $a=a_0$ having "thickness" $r_0$, and is
ghost, i.e. particle-twin in all other universes including our Universe 
($a=~0$).

At stage  ${\bf 1}=\ell C^\infty (\R^0)=\ell C^\infty (\R)/(a^1)$ \ \ \
$\theta\cdot f(x^3+x^0)mod \{a^1\}=f(x^3+x^0)$.  It means that we have
usual particle carryinging energy and momentum.

\section{Photon ghosts and photons-twins}

It is known that ftat monochromatic electro-magnetic wave
is described by wave equation
$$
\frac{1}{c}\frac{\partial \vec{\bf A}}{\partial t}=\Delta \vec{\bf A}
$$
and has, for example, the following form
$$
\vec{\bf A}=\vec{\bf A_0}e^{i(\vec{k}\vec{x}-\omega t)}.
$$
Electric and magnetic field strengthes of wave are equal to
\begin{equation}\label{e}
 \vec{\bf E}=i|\vec{k}|\vec{\bf A},\ \ \vec{\bf H}=i[\vec{k}\times
\vec{\bf A}].
\end{equation}
For energy-momentum tensor of wave we have
$$
T^{ij}=\frac{Wc^2}{\omega^2}k^ik^j,
$$
where
$$
W=\frac{\vec{\bf E}^2}{4\pi}
$$
is energy density of wave.

It follows from these formulas that under substitution
 $\vec{\bf A}\to d\vec{\bf A}$,
where $d\in D$, we can get
$$
\vec{\bf E}\to d\vec{\bf E} \Longrightarrow
\vec{\bf E}(\ell  C^\infty (\R)/(a^2))\neq 0 \ \mbox{under}\ a\neq 0.
$$
But $W\to d^2W=0$. Hence $T_{ik}\equiv 0$, i.e. we have photon ghost 
in all universes
of multiverse. This photon ghost is electro-magnetic wave
which is not carrying neither energy, nor momentum in all universes, 
except universe
with $a=0$, where it does not exist.

\medskip

Consider now a number $\vartheta \in R$. Let at stage
$\ell C^\infty (\R)/I$ it is defined by class functions 
$\vartheta(a)mod\ I$, where
\begin{equation}\label{f}
  \vartheta(a)=e^{-\gamma|a|^2}-1, \ \ \gamma>0.
\end{equation}
We get by means of substitution $\vec{\bf A}\to \vartheta\vec{\bf A}$ 
from (\ref{e}):
$$
\vec{\bf E}=i\vartheta|\vec{k}|\vec{\bf A},\ \ \vec{\bf H}=
i\vartheta[\vec{k}\times\vec{\bf A}],
 \ \ \vec{\bf
A}\neq 0.
$$

Then
$$
\vec{\bf E}(\ell C^\infty(\R)/(\vartheta^2))\neq 0,
$$
but
$$
T^{ij}=\frac{Wc^2}{\omega^2}k^ik^j(\ell C^\infty(\R)/(\vartheta^2))\ 
mod\ (\vartheta^2)=0.
$$

In other words
at stage (environment) $\ell C^\infty(\R)/(\vartheta^2)$ photons-twins
which are not carrying neither energy, nor momentum (i.e. being photon 
ghosts)
are  observed in all universes.

\section{Virtual reality as topos models
of formal multiverse }

"Set of real numbers" $R$ in ${\bf Set}^{\L^{op}}$ has no many
accustomed properties of real numbers from $\R$. Hence
existence in environments of
this virtual reality generator implies unexpected or unaccustomed 
facts and phenomena.
Some such facts were described in giving paper.

Topos ${\bf Set}^{\L^{op}}$ is not unique model for formal theory $\T$.
Other models, i.e. other virtual reality generators, will demonstrate
new properties, new realities. But it is difficult to say
which virtual reality is our own Physical Reality.



\begin{thebibliography}{99}

\bibitem{1}
    Deutsch, D. {\it The Fabric of Reality.} Allen Lane. 
The Penguin Press, 2000.


\bibitem{Cock}
Kock, A. {\it Synthetic Differential Geometry.} Cambridge Univ. 
Press, 1981.

\bibitem{gugri}
Guts, A.K., Grinkevich, E.B. {\it Toposes in General Theory of Relativity}.\\
 -- Los Alamos E-print
paper: gr-qc/9610073 (1996).\\ - http://xxx.lanl.gov/abs/gr-qc/9610073

\bibitem{gu}
Guts, A.K. {\it Intuitionistic theory of space-time} //
International geometric school-seminar in memory of
N.V.~Efimov. Abstracts. Abrau-Dyurso. September 27 - October 4, 
1996. P.87-88.

\bibitem{ged}
G\"odel, K. {\it An Example of a New Type of Cosmological Solution 
of Einstein's Field
Equations of Gravitation.} // Rev. Mod. Phys. 1949. V.21, No.3. P.447-450.

\bibitem{Dir}
Dirac, P. {\it Principles of Quantum Mechanics.} Moscow: Nauka, 1979.

\bibitem{Moer}
Moerdijk, I., Reyes, G.E. {\it Models for Smooth Infenitesimal Analysis.}
Springer-Verlag, 1991.

\bibitem{ever}
{\it Quantum Mechanics of Everett.}  -- Site in InterNet: \\
http://www.univer.omsk.su/omsk/Sci/Everett.
\bibitem{guz}
Guts, A.K., Zvyagintsev, A.A. {\it Interpretation of  intuitionistic 
solution of
the vacuum Einstein  equations in smooth topos}. -- Los
Alamos E-print Paper: gr-qc/0001076 (2000).

\bibitem{guz1}
Guts, A.K., Zvyagintsev, A.A.  {\it Solution of  nearly vacuum 
Einstein equations in
Synthetic Differential Geometry}
// Mathematical Structures and Modeling. 2000.
No.6. P.115-127.

\bibitem{guz2}
Guts, A.K., Zvyagintsev, A.A.
 {\it Intuitionistic Logic and Signature of
 Space-time}
// Logic and Applications. International Conference on the 60 birthday of
Yu.L.~Ershov. Abstracts. -- Novosibirsk: Institute of Discrete Math. 
and Informatics.
2000. P.38-39.


\bibitem{guz3}
Guts, A.K. {\it Many-valued Logic and multi-variant World} //
Logic and Applications. International Conference on the 60 birthday of
Yu.L.~Ershov. Abstracts. -- Novosibirsk: Institute of Discrete Math. 
and Informatics.
2000. P.36-37.


\bibitem{guz4}
Guts, A.K. {\it Interaction of the Past of parallel universes}.
- Los Alamos E-print Paper: physics/9910037 (1999).

\bibitem{guz5}
Guts, A.K.  {\it Models of multi-variant History} //
 Mathematical Structures and Modeling. 1999.
No.4. P.5-14.


\bibitem{Pal}
Palesheva E.V. {\it Ghost spinors, shadow electrons and the 
Deutsch Multiverse.}
-- Los Alamos E-print paper: gr-qc/0108017 (2001).

\end{thebibliography}
\end{document}